\documentclass[10pt]{iopart}

\usepackage{cite}
\usepackage{color}
\usepackage{graphicx}
\usepackage{iopams}

\expandafter\let\csname equation*\endcsname\relax

\expandafter\let\csname endequation*\endcsname\relax

\usepackage{amsmath}
\usepackage{amssymb}

\begin{document}

\title[Flexibility defines structure in crystals of amphiphilic DNA nanostars]
  {Flexibility defines structure in crystals of amphiphilic DNA nanostars}

\author{Ryan A. Brady$^1$, Will T. Kaufhold$^1$, Nicholas J. Brooks$^2$, Vito Foder\`{a}$^3$, and Lorenzo Di Michele$^1$}

\address{$^1$Biological and Soft Systems, Cavendish Laboratory, University of Cambridge, Cambridge CB3 0HE, UK}

\address{$^2$Department of Chemistry, Imperial College London, London SW7 2AZ, UK}

\address{$^3$Department of Pharmacy, University of Copenhagen, Universitetsparken 2
2100 Copenhagen, Denmark}

\ead{ld389@cam.ac.uk}

\vspace{10pt}
\begin{indented}
\item[]August 2017
\end{indented}

\begin{abstract}
DNA nanostructures with programmable shape and interactions can be used as building blocks for the self-assembly of crystalline materials with prescribed nanoscale features, holding a vast technological potential. Structural rigidity and  bond directionality have been recognised as key design features for DNA motifs to sustain long-range order in 3D, but the practical challenges associated with prescribing building-block geometry with sufficient accuracy have limited the variety of available designs. We have recently introduced a novel platform for the one-pot preparation of crystalline DNA frameworks supported by a combination of Watson-Crick base pairing and hydrophobic forces [\emph{Nano Lett.}, 17(5):3276--3281, 2017].
Here we use small angle X-ray scattering and coarse-grained molecular simulations to demonstrate that, as opposed to available all-DNA approaches, amphiphilic motifs do not rely on structural rigidity to support long-range order. Instead, the flexibility of amphiphilic DNA building-blocks is a crucial feature for successful crystallisation. 
\end{abstract}

%
\vspace{2pc}

%
\submitto{\JPCM}
%
\maketitle
%
\ioptwocol

\section{Introduction}

Developing methods to produce 3D crystalline arrays with arbitrary nanoscale structure is critical to a variety of emerging technologies including photonics,\cite{Flory2011} energy storage,\cite{Arico2005,Zhang2013} molecular filtration,\cite{Han2008} and sensing.\cite{Huang2007,Xia20101,Jimenez-Cadena} As a material to design and build at the nanoscale, DNA boasts numerous highly advantageous properties including high binding specificity, ease of functionalisation, prescribable interaction strength, and steadily decreasing production costs. Since the idea of structural DNA nanotechnology was introduced by Seeman \cite{Seeman1982,Seeman1985}, these unique properties have been exploited to build complex nanoscale structures of near-arbitrary shape~\cite{Seeman:2017aa,Jones2015}, and methods have been developed to propagate nanoscale structural control over macroscopic length-scales in one~\cite{Li2004,Rahbani:2015aa,Hariri:2015aa}, two~\cite{Winfree1998,Park2006,Hansen2010,Selmi2011,Lee2013,Avakyan:2017aa}, and three dimensions \cite{Zheng2009,Sha:2013aa,Zhang2018}.\\
Nearly all examples of DNA motifs capable of supporting long-range order in 3D rely on rigid building-blocks and on imposing the sought-after crystal structure through the formation of bonds with prescribed orientation~\cite{Jones2015}. This is the case for the stiff tensegrity triangles originally introduced by Zheng \emph{et al.}~\cite{Zheng2009,Sha:2013aa} and recently scaled up by Zhang \emph{et. al} using the DNA origami technique~\cite{Zhang2018}.\\
Although hypothetically straightforward, the use of rigid building blocks conceals practical challenges associated with the need for an extremely precise control over their 3D shape, which has so far hampered the development of a wide variety of designs.
Seeman's original proposal for the creation of 3D crystalline frameworks was based on much simpler motifs, namely branched DNA molecules\cite{Seeman1985}, or \emph{DNA nanostars}. These comprise of multiple double-helical arms radiating from a central point with each arm typically terminating with a single-stranded domain (\emph{sticky end}) to allow for specific interaction between units. Nanostars have become ubiquitous elements in the production of DNA architectures, from the formation of closed nanoscale objects~\cite{Chen1991,Seeman2003} to functional macroscopic DNA hydrogels\cite{Um2006,Park2009,Lee:2012aa,Hur2013}. The ability to tune the valency and strength of the interactions by prescribing the number of arms and the sequence of the sticky ends, also made nanostars ideal experimental models to unravel the phase behaviour of particles with valence-limited attractive interactions~\cite{Biffi2013,Rovigatti2014a,Rovigatti2014,Locatelli2017}.
However, conventional DNA nanostars have never been observed to form crystal phases. In fact,  computer simulations have demonstrated that crystallisation is thermodynamically forbidden owing to the excessive flexibility of these motifs~\cite{Rovigatti2014a}, consistent with the empirical finding that crystallisation of DNA nanostructures requires structural rigidity~\cite{Jones2015}.\\
\begin{figure}[t!]
\includegraphics[width=8cm]{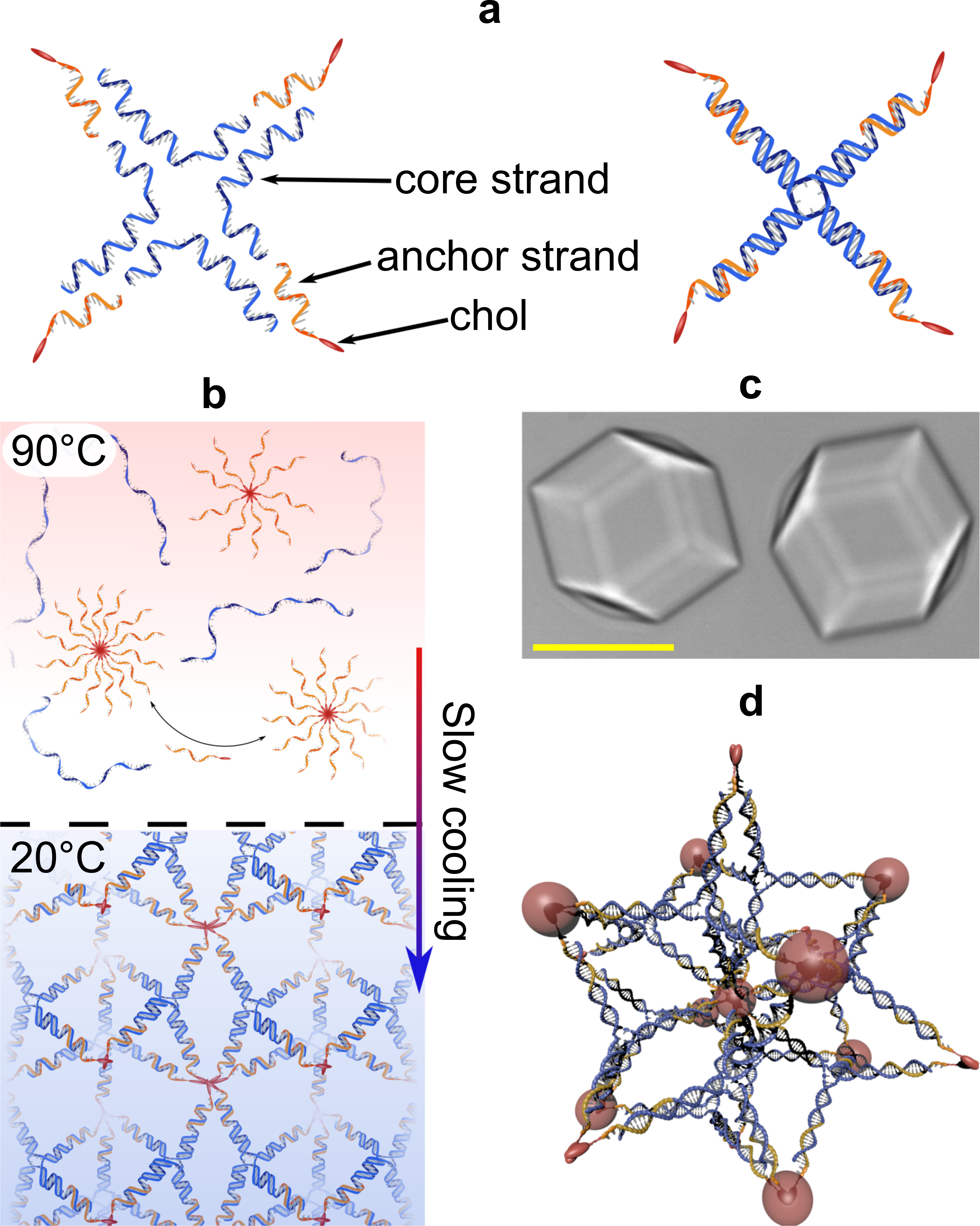}
\caption{\label{figure1} \textbf{Structure and self-assembly of amphiphilic C-stars.} \textbf{a}, Cholesterol functionalised DNA nanostars (\emph{C-stars}) with 4-arm topology self-assemble from 4-cholesterol modified strands (orange) and 4 non-functionalised core strands (blue)~\cite{Brady2017,Ryan2}. \textbf{b},~All single-stranded components are mixed in stoichiometric ratio, incubated at high temperature and then slowly cooled to form a network phase in which cholesterol-rich cores are cross-linked by the nanostar motifs. \textbf{c}, In suitable conditions, C-stars form macroscopic single crystals with BCC symmetry, highlighted here by bright-field micrographs~\cite{Brady2017,Ryan2}. Scale bars 20\,$\mu$m. \textbf{d},~Hypothesised arrangement of C-stars within the BCC unit cell, with lattice points highlighted by red circles.}
\end{figure}
Recently, we have demonstrated that by relaxing the fixed-valency constraint, flexible DNA junctions can in fact crystallise~\cite{Brady2017,Ryan2}. We have achieved this by replacing the specifically-binding DNA sticky ends at the end of each arm with hydrophobic cholesterol molecules, making interactions non specific and multivalent as for amphiphilic star polymers~\cite{Capone2012,Koch2013} (Fig.~\ref{figure1}). Despite the lack of binding specificity, the phase behaviour of the amphiphilic DNA nanostars, or \emph{C-stars}, can still be controlled by changing the number of arms~\cite{Brady2017}, a feature much simpler to prescribe than the rigid 3D geometry one needs to control with all-DNA building blocks \cite{Zheng2009,Sha:2013aa,Zhang2018}.\\
Here, through a combination of molecular dynamics simulations and small angle X-ray scattering, we dig deeper into the role that structural rigidity plays in C-star crystallisation. We tune the flexibility of four-arm C-stars by  symmetrically including unpaired bases at the central junction, and further by using buffers containing divalent rather than monovalent cations. We demonstrate how seemingly minor changes in nanostructure or buffer conditions lead to substantial differences in the structure of the network phases, and show how flexibility is not only an acceptable characteristic, but in fact a critical feature for successful crystallisation of amphiphilic DNA motifs. \\

\section{Results and discussion}

\subsection{C-star design and flexibility}

As shown in Fig.~\ref{figure1}\textbf{a}, 4-arm C-stars relevant to this work are composed of 4 \emph{core} single-stranded (ss) DNA molecules designed to form a tetravalent junction (blue) and 4 cholesterol-functionalised ssDNA molecules that connecting to the central junction result in the presence of a hydrophobic tag at the end of each arm (orange), making C-stars amphiphilic~\cite{Brady2017}. For self-assembly, the single-stranded components are mixed in stoichiometric ratio, then heated up above the melting temperature of all duplexes (95$^\circ$C) and slowly cooled down to room temperature (20$^\circ$C) at a rate of -0.01$^\circ$C min$^{-1}$. At high temperature, individual core strands coexist with cholesterol-DNA micelles. As the temperature is decreased, nanostar motifs form, and start cross-linking the micelles, until a phase transition is encountered and extended framework phases are formed~\cite{Brady2017} (Fig.~\ref{figure1}\textbf{b}). Under most experimental conditions, these frameworks display long-range crystalline order and form macroscopic single crystals exceeding 50\,$\mu$m in size~\cite{Brady2017,Ryan2} (Fig.~\ref{figure1}\textbf{c}).\\
Cholesterol-tagged strands bind to the core nanostar motif through a 14-base overhang. At room temperature, the free energy gain associated with the formation of such bond is $\sim - 45\,k_\mathrm{B}T$~\cite{Markham:2005aa,SantaLucia1460}. In turn, the free energy loss associated to extracting a cholesterol moiety from a micelle is estimated in  $\sim 16\,k_\mathrm{B}T$, calculated assuming a critical micellar concentration of $160$\,nM~\cite{Choi:2013aa}. It is thus safe to assume that, at equilibrium, all C-star motifs are well formed and have the prescribed 4-arm topology. Given this hard constraint, the equilibrium structure of the frameworks is then controlled by the micelle-like cores adapting their coordination and, most importantly for the purpose of this work, the ability of the DNA mortifs to adapt their geometry by flexing.\\
Configurational flexibility can arise in C-stars through bending of the individual arms or pivoting at the central junction. Each arm features a nick half-way along (Fig.~\ref{figure1}), which may facilitate arm bending. However, stacking interactions are known to strongly stabilise a continuous double-helical structure at nick sites, as confirmed by dedicated experiments~\cite{Roll:1998aa,Zhang:2003aa,Kashida:2017aa} and by \emph{in silico} observations discussed below.  The bending of the double-helical arms irrespective of the nick is also expected to have a minor contribution on the flexibility of the nanostar motifs, as the persistence length of double-stranded DNA exceeds the overall arm-length by a factor $\gtrsim$4~\cite{Manning:2006aa}.
We can thus safely identify the central junction of the nanostars as the main source of their flexibility, which in this contribution we fine tune by varying the number of free bases between neighbouring arms, $N_\mathrm{free}$. Specifically, we test C-star variants with $N_\mathrm{free}$ of 0 and 4, to compare a stiffer and a more flexible design with the  previously investigated variant with $N_\mathrm{free}=1$~\cite{Brady2017,Ryan2} (Fig.~\ref{figure3}\textbf{a}). \\
\begin{figure}[t!]
\includegraphics[width=8cm]{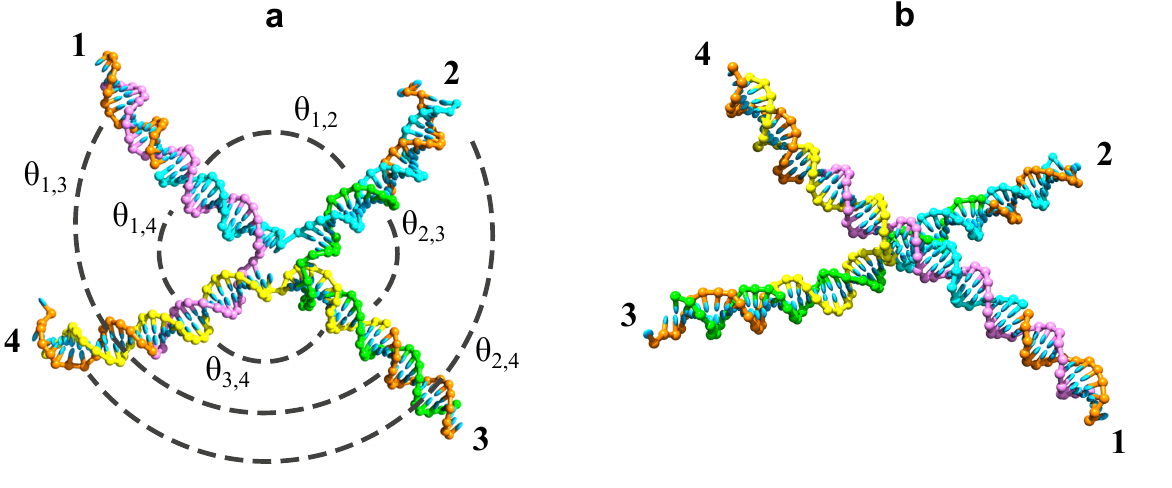}
\caption{\label{figure2} \textbf{Geometry of a 4-arm DNA nanostar.}  \textbf{a},~Definition of arms and inter-arm angles. \textbf{b},~A stacked X-structure. Images of nanostars are snapshots of coarse-grained molecular simulations~\cite{oxDNA,oxDNA2} rendered using UCSF Chimera~\cite{chimera}.}
\end{figure} 
\begin{figure*}[ht!]
\includegraphics[width=15.96cm]{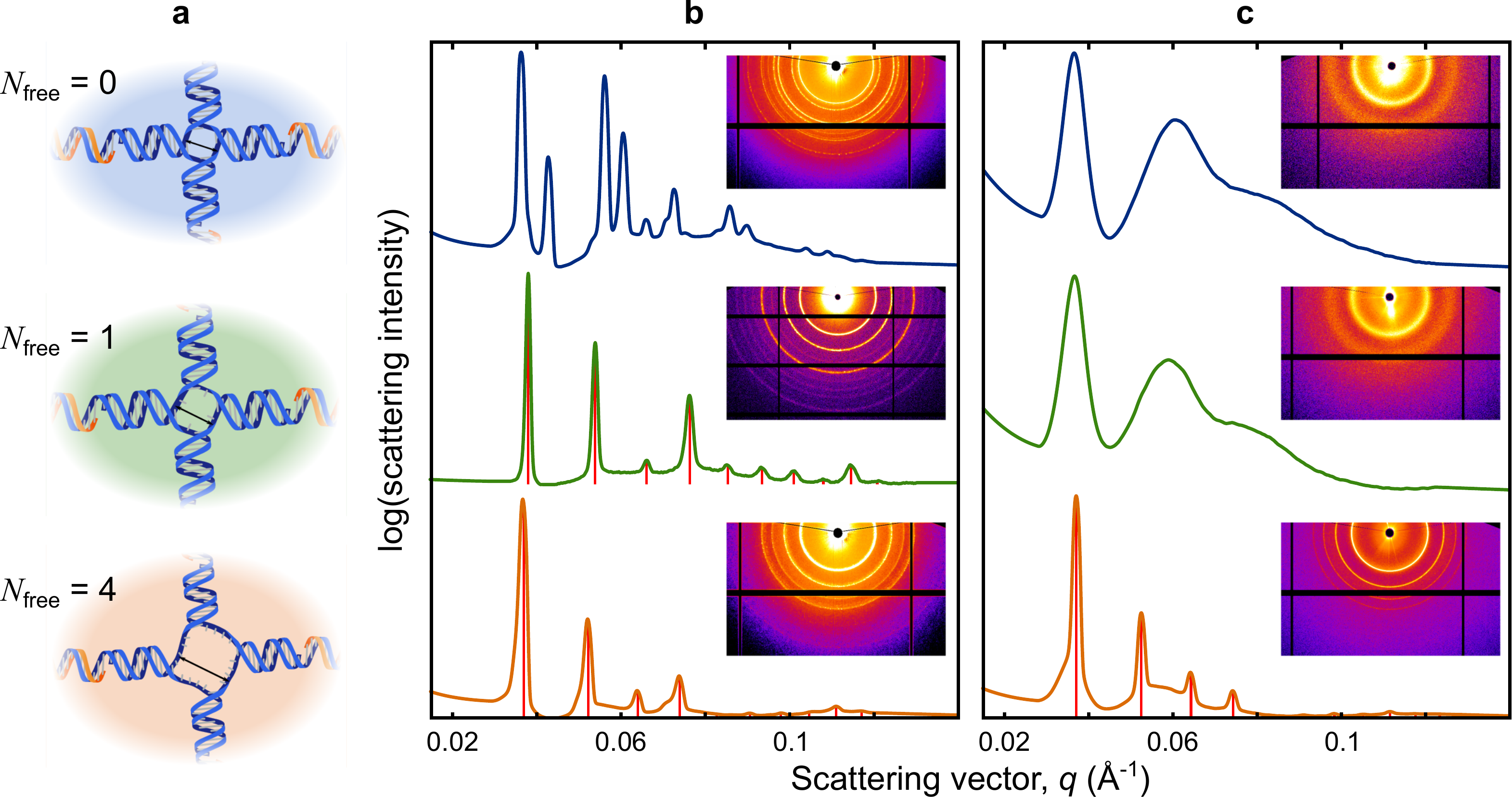}
\caption{\label{figure3} \textbf{Crystal structure of C-star networks depends on the number of free bases at the central junction.} \textbf{a},~Detail of the unpaired bases at the junctions of the three tested C-star designs with $N_\mathrm{free}=0,1$ and $4$. \textbf{b-c}, 2D (insets) and radially averaged SAXS profiles of C-stars networks self-assembled in TE buffer supplemented with either 300\,mM NaCl (\textbf{b}) or 17\,mM MgCl$_{2}$ (\textbf{c}). Where present, red vertical lines mark the best fit to the Bragg peaks of a BCC phase, which for $N_\mathrm{free}=1$ and $N_\mathrm{free}=4$ result in a lattice parameter $a=232$\,\AA~and $a=240$\,\AA~respectively.}
\end{figure*}
With reference to Fig.~\ref{figure2}\textbf{a}, if we label each C-Star arms as 1$\dots$4, the conformation of the junction is defined by six angles $\theta_{ij}$ between 0 and 180$^\circ$ ($i,j = 1\dots4, i\ne j$). Pairs of arms are defined as adjacent if they share a strand, and opposite if they do not, so that each arm is adjacent to two others and opposite to the third. In Fig.~\ref{figure2}, for instance, arm 1 is adjacent to arm 2 and arm 4, and opposite to arm 3.  Four of the six angles, therefore, are defined between adjacent arms, and the remaining two between opposite arms. In Fig.~\ref{figure2}, $\theta_{1,2}, \theta_{2,3}, \theta_{3,4}$ and $\theta_{1,4}$ are \emph{adjacent angles}, while $\theta_{1,3}$ and $\theta_{2,4}$ are \emph{opposite angles}.\\
Fully base-paired junctions ($N_\mathrm{free}=0$) are known to adopt stiff X-shaped configurations for sufficiently high ionic strength, in which each arm stacks to one adjacent neighbour, and  two quasicontinuous helices are formed\cite{Duckett1990,lilley1993,DUCKETT198879,Cooper1989,ALTONA1996305}, as highlighted in Fig.~\ref{figure2}\textbf{b}. In this configuration two of the adjacent angles are $\sim$180$^{\circ}$, the other two adjacent angles are $<90^{\circ}$, and the opposite angles $>90^{\circ}$.
Inclusion of unpaired bases has been shown to destabilise the stiff X-structure and cause junctions to adopt a more compliant square planar conformation with $\sim90^{\circ}$ angles between adjacent arms and $\sim180^{\circ}$ between opposite arms, or a flexible tetrahedral geometry where arms can freely pivot around the central junction and all inter-arm angles average 109.5$^{\circ}$\cite{DUCKETT1991147}.\\
The geometry of branched DNA is also highly dependent on the valency and concentration of counterions present. High ionic strength stabilises the stacked configuration, and divalent ions are much more effective in doing so than monovalent ions at the same ionic strength~\cite{Cooper1989,Duckett1990,Lilley4Junction,Nollmann2004}. To investigate the effect of cation identity on the structure of amphiphilic DNA crystals, we also compare aggregates grown in the presence of NaCl with those grown in buffer supplemented with MgCl$_{2}$.\\

\begin{figure*}[t!]
\includegraphics[width=17cm]{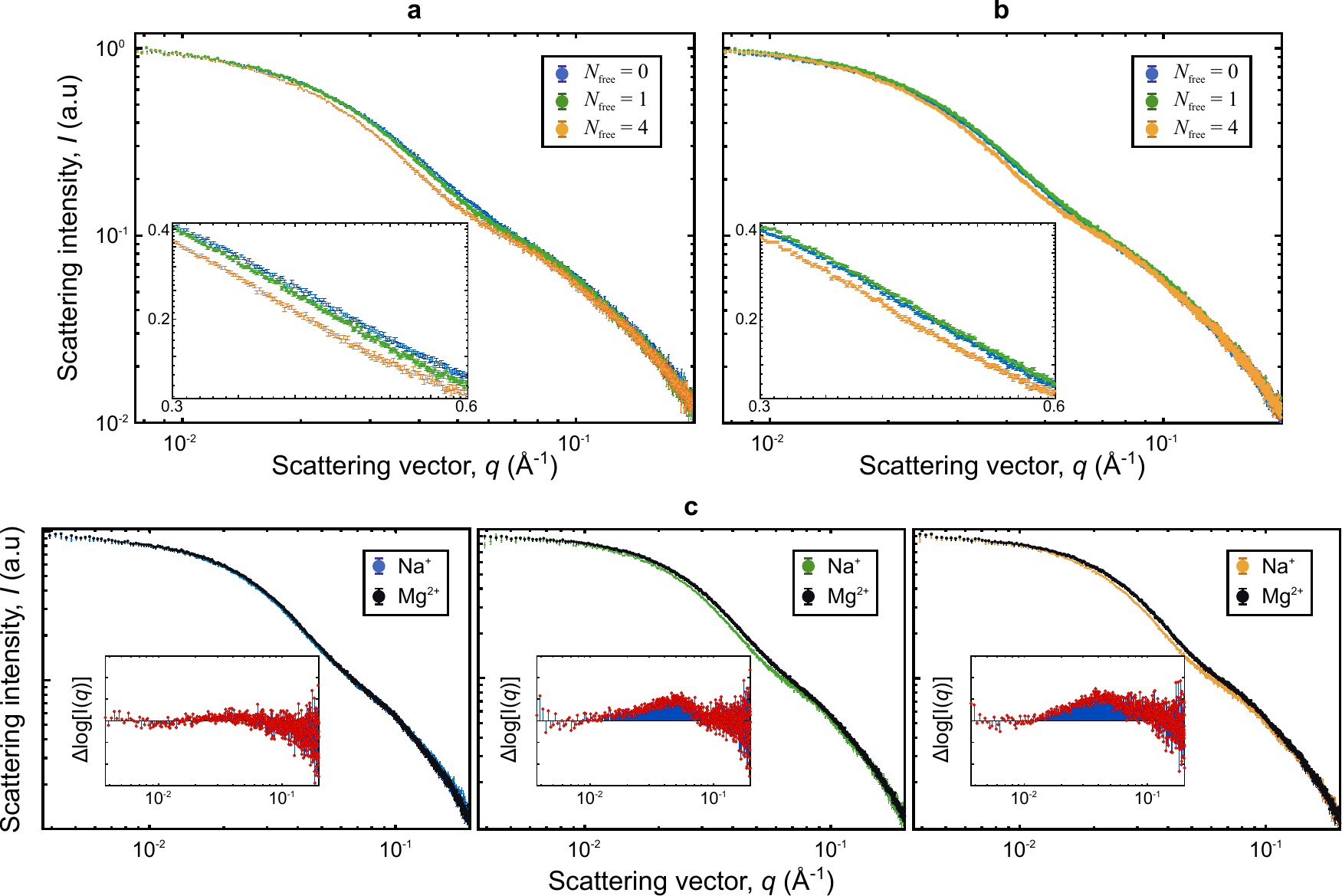}
\caption{\label{figure4} \textbf{Solution-SAXS highlights structural differences between nanostar designs and buffer conditions.}  Scattering intensity of the three nanostar designs with $N_\mathrm{free}=0,$ 1 and 4 carried out in 300\,mM NaCl (\textbf{a}) and 17\,mM MgCl$_{2}$ (\textbf{b}). \textbf{c}, Direct comparison of SAXS traces of each nanostar design in different buffer conditions. Insets show the difference between the logarithms of the scattering intensities measured in 300\,mM NaCl and 17\,$\mu$M MgCl$_{2}$.}
\end{figure*}

\subsection{Effect of flexibility on long range order}

We have previously shown that samples of 4-arm C-stars with $N_\mathrm{free}=1$ prepared in buffer with 300\,mM NaCl crystallise with a body-centre-cubic (BCC) symmetry~\cite{Brady2017,Ryan2}. This is demonstrated in Fig.~\ref{figure3}\textbf{b} (centre), where we show a radially-averaged diffraction pattern collected by synchrotron-based small-angle X-ray scattering (SAXS) of ``powder'' samples (see Methods). The Bragg peaks are perfectly consistent with a BCC symmetry and lattice parameter $a=232$\,\AA~\cite{Ryan2}, but a direct determination of the distribution of the building blocks within the unit cell has so far proven elusive. Nonetheless, based on the dependence of the lattice parameter on C-star arm-length, the measured porosity of the frameworks, and the expected coordination of the DNA-cholesterol micelles, we hypothesise an arrangement with 6 C-stars per unit cell cross-linking cholesterol-rich cores positioned at the BCC lattice points, where 12 C-star arms converge~\cite{Brady2017,Ryan2} (Fig.~\ref{figure1}\textbf{d}). Within the hypothesised unit cell, C-stars assume a quasi-tetrahedral geometry, with four of the inter-arm angles equal to 102$^{\circ}$ and the remaining two angles equal to 127$^{\circ}$.\\
In Fig.~\ref{figure3}\textbf{b} we compare SAXS diffraction patterns of C-star samples featuring $N_\mathrm{free}=0$ and $N_\mathrm{free}=4$, with the previously investigated design with $N_\mathrm{free}=1$, all prepared in buffers with 300\,mM NaCl. The nominally most flexible variant, with $N_\mathrm{free}=4$, adopts a BCC phase identical to the design with $N_\mathrm{free}=1$, but featuring a slightly larger lattice parameter $a=240$\,\AA. Assuming that the arrangement of C-stars within the unit cell is conserved (Fig.~\ref{figure1}\textbf{d}), this slight increase can be readily explained with a comparatively expanded central junction of the $N_\mathrm{free}=4$ variant due to the additional unpaired bases. In contrast to the other two designs, the stiffest variant trialled with $N_\mathrm{free}=0$ forms a lower symmetry (non-cubic) lattice, or possibly features multiple coexisting phases.\\
 Figure~\ref{figure3}\textbf{c} shows the SAXS patterns of network phases grown in the presence of magnesium for the three C-star designs. In contrast to samples prepared in sodium, here only the variant with $N_\mathrm{free}=4$ forms a crystalline phase, again adopting a BCC lattice with $a=240$\,\AA, while both the $N_\mathrm{free}=0$ and $N_\mathrm{free}=1$ appear to form hydrogels with two coordination shells.\\
\begin{figure*}[ht!]
\begin{center}
\includegraphics[width=14.67cm]{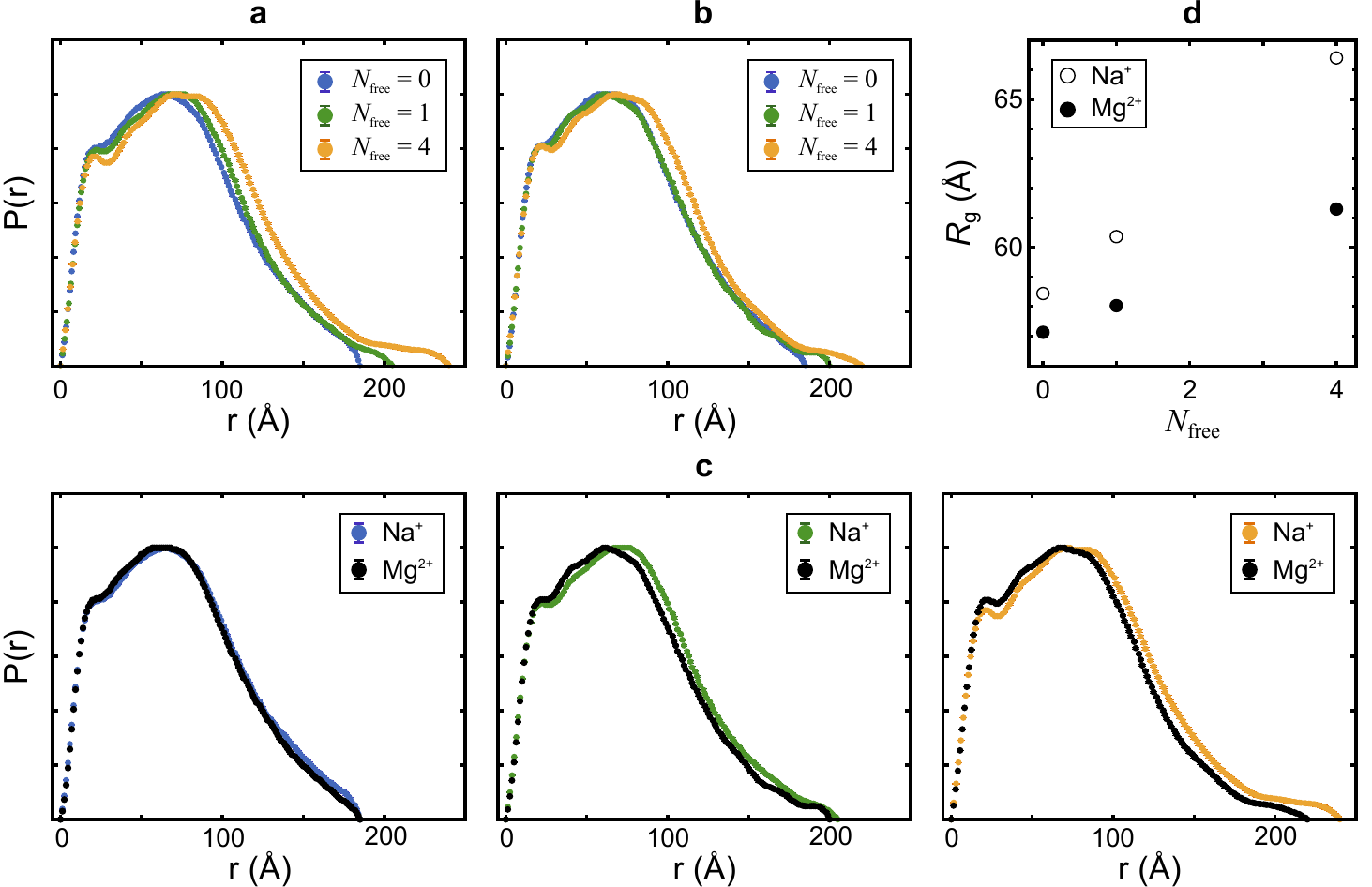}
\caption{\label{figure5} \textbf{Internal pair-distribution function and radius of gyration of individual nanostar highlight structural differences between different designs and buffer conditions.}  Internal pair distribution function $P(r)$ determined using the software GNOM for nanostar variants prepared in 300\,mM NaCl (\textbf{a}) and 17\,mM MgCl$_{2}$ (\textbf{b}). \textbf{c}, Direct comparison of $P(r)$ for each nanostar between buffer conditions tested. \textbf{d}, Radii of gyration, $R_{\mathrm{g}}$ as extracted by Guinier analysis of the experimental scattering curves.}
\end{center}
\end{figure*}
The observation that highly flexible designs with $N_\mathrm{free}=4$ are always capable of forming the high-symmetry BCC phase, while the stiffest junctions with $N_\mathrm{free}=0$ can never do so, regardless on cation identity, demonstrates that conformational flexibility is indeed a critical factor enabling C-star crystallisation. This deduction is further corroborated by the disruptive effect of magnesium, which is known to stabilise stiff stacked conformations~\cite{Duckett1990,lilley1993,DUCKETT198879,Cooper1989,ALTONA1996305,JOO2004739} and completely prevents crystallisation for $N_\mathrm{free}=1$ and $N_\mathrm{free}=0$.\\
To further rationalise the effect of junction stiffness  on crystallisation, in the reminder of this paper we use a combination of solution-based SAXS and coarse-grained molecular dynamic simulations to investigate equilibrium configuration and flexibility of individual nanostars in conditions relevant to the crystallisation experiments.\\

\subsection{Conformation of individual DNA nanostars}
Solution-based SAXS is used to characterise the morphology of non-interacting nanostars lacking the cholesterol moieties, but otherwise identical to the C-stars used for crystallisation experiments. 
For samples prepared in  300\,mM NaCl,  the scattering intensity $I(q)$ shows a clear trend in the intermediate range of the scattering vector $q$, $0.02<q<0.08$\,\AA$^{-1}$, as demonstrated in Fig.~\ref{figure4}\textbf{a} and further highlighted in the inset. We observe that, as $N_\mathrm{free}$ is increased, the characteristic ``bump'' in the scattering trace becomes more pronounced and shifts towards lower $q$. Consistently, similar trends have been observed for increasing salt concentration, and have been linked to expansion of the central junction and an increasing ability of the arms to fluctuate~\cite{SANSFree}.
Figure~\ref{figure5}\textbf{a} shows the internal pair distribution functions $P(r)$ of the nanostars, extracted from $I(q)$ using the indirect transform program GNOM\cite{Svergun:wi0087}. The progressive shift towards greater distances of the decaying edge of $P(r)$, observed as $N_\mathrm{free}$ is increased, confirms the expansion of the nanostructure as a whole, which is further demonstrated by the increase in the radius of gyration $R_\mathrm{g}$ calculated through Guinier analysis~\cite{Guinier} (Fig.~\ref{figure5}\textbf{d}). The overall larger size of the motif with $N_\mathrm{free}=4$ compared to $N_\mathrm{free}=1$ is consistent with the increase in lattice parameter of crystals produced form these motifs, as shown in Fig.~\ref{figure3}.\\
For samples prepared in 17\,mM MgCl$_{2}$ the variations in $I(q)$ and $P(r)$ associated to different $N_\mathrm{free}$ are significantly smaller in comparison with the ones observed with sodiuim, as demonstrated in Fig.~\ref{figure4}\textbf{b} and Fig.~\ref{figure5}\textbf{b}, respectively. In particular, designs with $N_\mathrm{free}=0$ and 1 prepared in magnesium show negligible differences.\\
Scattering traces collected for the same $N_\mathrm{free}$ in the two different salts are directly compared in Fig.~\ref{figure4}\textbf{c}, while the same comparison is made between the pair distribution functions in Fig.~\ref{figure5}\textbf{c}. For designs with $N_\mathrm{free}=1$ and $N_\mathrm{free}=4$, magnesium causes the characteristic bump to flatten out and shift to higher intensity, which corresponds to a reduction in overall nanostar size demonstrated by a shift in the $P(r)$ and more directly by a significant decrease in $R_g$ (Fig.~\ref{figure5}\textbf{d}). In turn, for $N_\mathrm{free}=0$, replacing sodium with magnesium appears to have a negligible effect on both $I(q)$ (Fig.~\ref{figure4}\textbf{c}) and $P(r)$ (Fig.~\ref{figure5}\textbf{c}), while producing only a marginal decrease in $R_g$ (Fig.~\ref{figure5}\textbf{d}).\\
Following previous studies\cite{Duckett1990,lilley1993,DUCKETT198879,Cooper1989,ALTONA1996305,JOO2004739}, we expect that DNA junctions without free bases adopt the fully stacked X-structure in the presence of magnesium or a relatively high concentration of sodium.  This is fully consistent with the near-identical scattering traces measured for non-interacting C-stars with  $N_\mathrm{free}=0$ in both 300 mM NaCl and 17 mM MgCl$_2$, which also exhibit  the most compact geometry among the ones observed.
One may thus interpret the failure of  C-stars with $N_\mathrm{free}=0$ to form the high-symmetry BCC phase as a sign of the incompatibility between the morphology that the motifs assume within it and the stacked X-structure. At first, this interpretation may appear to clash with the observed difference between the microstructures of frameworks prepared in sodium and magnesium, with the former displaying a low-symmetry crystalline phase and the latter forming a hydrogel (Fig.~\ref{figure3}\textbf{b,c}).  However, it has been observed that for $N_\mathrm{free}=0$ the stability of the stacked X-structure is impacted by the identity of the counterions, with Mg$^{2+}$ shown to significantly reduce the rate of conformational change in comparison to Na$^{+}$~\cite{JOO2004739}. It is therefore plausible that Mg$^{2+}$ fully stabilises the stacked X-structure in C-stars with $N_\mathrm{free}=0$, preventing crystallisation altogether, but in the presence of sodium the flexibility of the junction is sufficient to enable a rearrangement compatible with the low-symmetry lattice observed.\\
For $N_\mathrm{free}=1$, the difference in the solution-scattering patterns measured in sodium and magnesium is ascribed to the formation, in the latter case, of the compact X-structure. This is consistent with the indistinguishability between scattering traces measured in MgCl$_2$ for $N_\mathrm{free}=0$ and $N_\mathrm{free}=1$ and with previous experiments\cite{DUCKETT1991147}, and explains the inability of C-stars with $N_\mathrm{free}=1$ to crystallise in magnesium (Fig.~\ref{figure3}\textbf{b}).
The comparatively expanded junction detected  for $N_\mathrm{free}=1$ in the presence of sodium hints at a greater flexibility that, we can argue, allows the nanostar motif to adapt to the BCC unit cell observe in C-star frameworks (Fig.~\ref{figure3}\textbf{a}).\\
The differences observed between the scattering profiles of nanostars with $N_\mathrm{free}=4$ and $N_\mathrm{free}=0,1$ (Fig.~\ref{figure4}\textbf{b}) demonstrate that, even in magnesium, the stacked X-structure is not stable if 4 free bases are left at the junction. The nanostars are thus expected to adopt a flexible tetrahedral (or square-planar) geometry, once again compatible with the high-symmetry BCC phase observed for C-star frameworks with $N_\mathrm{free}=4$ in both sodium and magnesium (Fig.~\ref{figure3}\textbf{b,c}).

\subsection{Conformational flexibility revealed by simulations}

To gain more direct insights on the likely conformation and flexibility of each nanostar variant, we perform molecular simulations of non-interacting nanostars using the coarse-grained model oxDNA2~\cite{oxDNA,oxDNA2}, as discussed in the Methods section. In the original oxDNA model, each nucleotide is represented by a rigid body interacting with its nearest neighbours \emph{via} irreversible backbone bonds, base stacking and excluded volume, and with all others nucleotides \emph{via} hydrogen bonds, coaxial and cross-stacking, and excluded volume~\cite{oxDNA}. Further to this, in oxDNA2, screened electrostatic interactions are implemented explicitly, and modelled by Yukawa potentials~\cite{oxDNA2}.\\\begin{figure*}[ht!]
\begin{center}
\includegraphics[width=14.95cm]{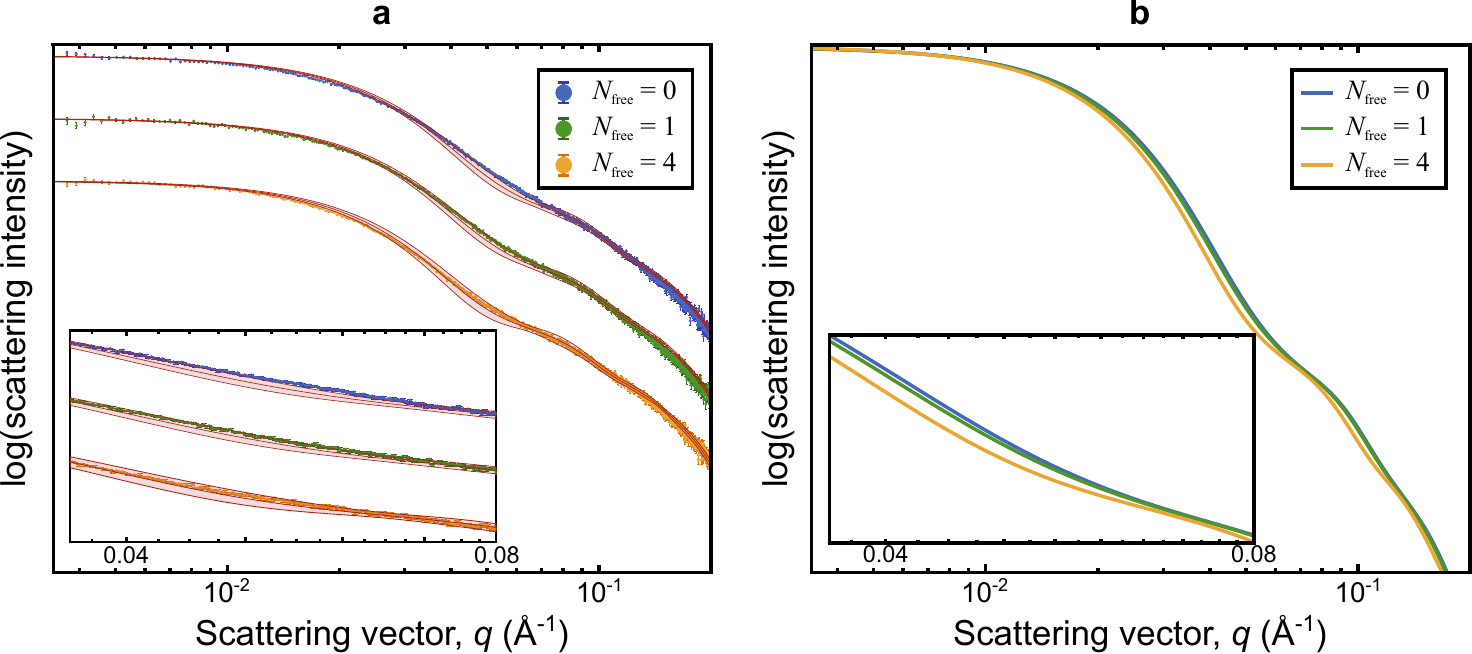}
\caption{\label{figure6} \textbf{Coarse-grained computer simulations reproduce experimental scattering traces}. \textbf{a}, Simulated scattering traces are marked by a red solid line and a shaded region, the former corresponding to the ensemble average calculated from simulated configurations and the latter defined by the standard deviation. Experiments for the  nanosar designs with $N_\mathrm{free}=0,$ 1, 4, performed in 300 mM NaCl, are overlaid to the corresponding simulated data. Curves corresponding to different samples are shifted by an arbitrary factor for clarity. \textbf{b}, Direct comparison between the simulated traces of panel \textbf{a}.}
\end{center}
\end{figure*}
As an initial validation of the simulation procedure, in Fig.~\ref{figure6}\textbf{a} we compare experimental solution-SAXS traces with simulated scattering profiles, as determined following the protocol introduced by Fernandez-Castanon \emph{et al.}~\cite{SANSFree} (see Methods). Since oxDNA2 does not discriminate based on ionic identity, simulations are performed at ionic strength equivalent to 300\,mM NaCl, and comparison in carried out with experimental data collected in the same conditions. No fitting steps are involved in comparing simulation and experiment besides normalisation by the intensity of the first 10 $q$ points. Simulated and experimental traces for nanostar designs with $N_\mathrm{free}=0, 1, 4$ are in good agreement over a wide $q$ range ($q<0.2$\,\AA). Consistently, simulated scattering curves reproduce the observed trend for increasing  $N_{\mathrm{free}}$, replicating the decrease in scattering intensity in the intermediate $q$ window $0.02<q<0.08$, as highlighted in Fig.~\ref{figure6}\textbf{b} and its inset. At higher values of $q$, simulated profiles show a second shoulder and differences between the three designs not evident in experimental traces, possibly masked by experimental noise.\\
Having observed quantitative agreement between simulated and experimental scattering traces, we further analyse oxDNA2 trajectories for deeper insights on the morphology of the nanostar motifs. As expected, simulations confirm that the nicks present half-way along the arms of the nanostars have nearly no effect on the flexibility of the structures, as coaxial stacking is found to stabilise a continuous double helix with 99.91\% probability. For each of the different nanostar designs we then sample the probability distributions of inter-arm angles $\theta_{i,j}$ and the cross-correlation between all pairs of angles, summarised in Fig.~\ref{figure7}.\\
For $N_\mathrm{free}=0$, the probability distributions of adjacent angles $\theta_{1,2}, \theta_{2,3}, \theta_{3,4}$ and $\theta_{1,4}$ show a clear bimodal pattern, with a sharp peak approaching 180$^\circ$ and a broader peak at $\theta_{i,j} \lesssim 90^\circ$, as expected for a stacked X-structure (Fig.~\ref{figure7}\textbf{a}). Consistently, the probability distribution of both opposite angles, $\theta_{1,3}$ and  $\theta_{2,4}$, features a single peak at $\theta_{i,j} \gtrsim 90^\circ$. The fingerprint of the stacked X-structure can be further identified in the two-angle probability maps. Here we observe strong negative correlation between adjacent angles relative to the same arm (\emph{e.g.} $\theta_{1,2}$ \emph{vs} $\theta_{1,4}$), demonstrating that when  one angle takes the high value associated with stacking, the other takes the smaller value corresponding to the bent configuration. Bimodality observed in these maps follows from the presence of two conformational isomers of the stacked X-structure, one in which arm 1 stacks to arm 2 and arm 3 stacks to arm 4, and the other in which arm 1 stacks to arm 4 and arm 2 stacks to arm 3. Isomerism also emerges from from the joint-probability maps between adjacent angles relative to different arms (\emph{e.g.} $\theta_{1,2}$ \emph{vs} $\theta_{3,4}$) that, as expected, also show a positive correlation. Note that oxDNA2~\cite{SANSFree}, similar to other coarse-grained models of DNA~\cite{Wang:2016aa}, does not to capture the correct handedness of stacked junctions, producing a left-handed geometry rather than the right-handed one determined experimentally~\cite{McKinney:2002aa}.\\
For junctions with $N_\mathrm{free}=1$,  probability distributions of the inter-arm angles and their cross-correlation maps show qualitatively similar trends to those computed for $N_\mathrm{free}=0$, suggesting that the X-structure remains the most stable configuration in free nanostars (Fig.~\ref{figure7}\textbf{b}). However, upon closer inspection, it is evident that for $N_\mathrm{free}=1$, peaks associated with the X-structure appear less pronounced than for the $N_\mathrm{free}=0$ design, hinting at a greater prominence of un-stacked, flexible, conformations. This is particularly evident in the probability distributions of adjacent angles (\emph{e.g.} $\theta_{1,2}$), where the two peaks are no longer resolvable, and in the correlation maps between adjacent angles relative to the same arm (\emph{e.g.} $\theta_{1,2}$ \emph{vs} $\theta_{1,4}$), where the region bridging the peaks corresponding to the two conformers appears to be significantly more populated.\\
\begin{figure*}[t!]
\includegraphics[width=17cm]{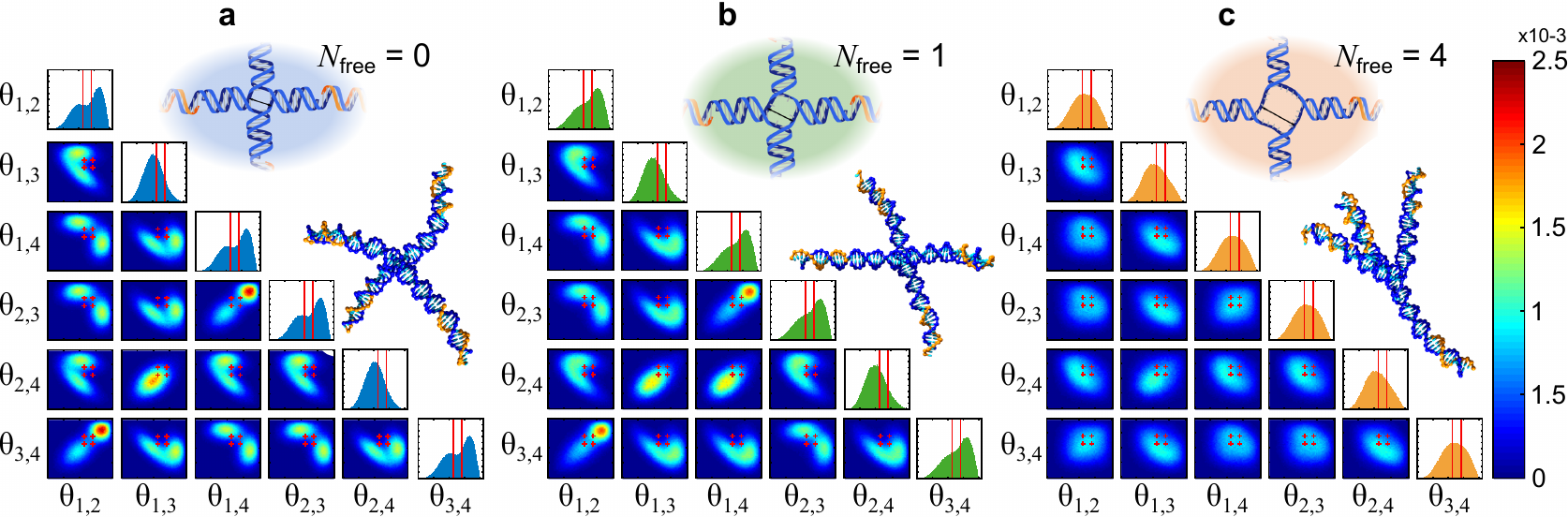}
\caption{\label{figure7} \textbf{Conformational flexibility of individual nanostars investigated by coarse-grained simulations.} For the three nanostar designs with  $N_\mathrm{free}=0$ (\textbf{a}), $N_\mathrm{free}=1$ (\textbf{b}) and  $N_\mathrm{free}=4$ (\textbf{c}), we show the combined probability distributions between pairs of inter-arm angles  (colour maps) and the probability distributions of individual angles (histograms), as computed using oxDNA2. Red symbols on the colour maps and red vertical lines on the histograms show the values of the inter-arm angles in the quasi-tetrahedral geometry assumed by the nanostars in the hypothesised BCC unit cell. For each panel, we show a typical simulation snapshot.}
\end{figure*}
The same data computed for $N_\mathrm{free}=4$  summarised in Fig.~\ref{figure7}\textbf{c}, highlight full conformational flexibility. The differences between the probability distributions of adjacent and opposite angles are minor, and all show a single broad peak compatible with the value expected for a flexible tetrahedron. Consistently, two-angle maps appear rather featureless, failing to highlight any strong correlation.\\
To qualitatively assess the likelihood of each nanostar design to assume the quasi-tetrahedral configuration required for the hypothesised arrangement of C-stars in the BCC unit cell, we highlight the relevant inter-arm angles (Fig.~\ref{figure2}) in single-angle distribution probabilities (red vertical lines) and the two-angle correlation maps (red symbols). Although when looking at the distributions of individual angles the probability of them matching the required values appears non-negligible, when accounting for two-angle correlations it is immediately clear that the stacked X-structure prominent for $N_\mathrm{free}=0$ and (less so) for  $N_\mathrm{free}=1$ is incompatible with the desired configuration. This is evident, for instance, in the correlation maps between adjacent angles relative to the same arm (\emph{e.g.} $\theta_{1,2}$ \emph{vs} $\theta_{1,4}$). 
In turn, for $N_\mathrm{free}=4$, the broad peaks comfortably encompass all the possible combinations of inter-arm angles satisfying the constraints of the BCC lattice.\\ 
Energy is required to distort C-stars away from their ground-state configuration in non-interacting conditions, which in dense phases is provided by the formation of a more or less optimal network of hydrophobic bonds. Regardless of whether the actual arrangement of C-stars in the BCC crystals corresponds to our hypothesis, simulations indicate that such a free energy cost is most significant for $N_\mathrm{free}=0$, slightly less prohibitive for $N_\mathrm{free}=1$, and smallest for $N_\mathrm{free}=4$. This is consistent with experimental trends on the emergence of BCC crystalline phases, which never occurs for $N_\mathrm{free}=0$ and always does for $N_\mathrm{free}=4$, and only emerges for $N_\mathrm{free}=1$ if Na$^+$ is used instead of Mg$^{2+}$ (Fig.~\ref{figure3}). The impossibility to distinguish between cations of different valency in simulations does not allow us to unravel this last effect.\\

\section{Conclusions}
In summary, through a combination of X-ray diffraction, solution scattering, and coarse-grained molecular simulations~\cite{oxDNA2}, we investigated the influence of conformational flexibility on the ability of recently introduced amphiphilic DNA nanostars to support long-range order in three dimensions~\cite{Brady2017,Ryan2}. Flexibility is systematically tuned by producing three nanostar designs with different number of unpaired bases at the central junction ($N_\mathrm{free}=0, 1, 4$), and by preparing samples in buffers supplemented with either monovalent (Na$^+$) or divalent (Mg$^{2+}$) cations.\\
Regardless of cation valency, C-stars with $N_\mathrm{free}=4$ form body-centred cubic crystal phases. The same crystal structure emerges in sodium buffers for C-star designs with $N_\mathrm{free}=1$, that however fail to crystallise in the presence of magnesium, forming instead an amorphous gel. Networks self-assembled from building blocks with $N_\mathrm{free}=0$ remain amorphous if grown with divalent ions, but exhibit a low-symmetry (non cubic) crystal phase if incubated in sodium.\\
We interpret these trends as a result of variations in the geometry and flexibility of the individual nanostar motifs. Junctions with $N_\mathrm{free}=4$ are highly flexible regardless on the nature of the cations, as confirmed by solution-based scattering and molecular simulations. Conformational flexibility allows the motifs to easily adapt to the conformation required for BCC crystallisation.  For $N_\mathrm{free}=1$ the use of magnesium results in junctions with a  rigid X-shape geometry, which we argue is incompatible with the BCC unit cell or any periodic lattice in 3D, hence hindering crytallisation. Scattering reveals that building blocks with the same design retain a greater flexibility in the sodium buffer and are arguably capable to relax into the geometry required for BCC crystallisation as for motifs with $N_\mathrm{free}=4$. The fully-stacked X-structure is observed in freely-suspended nano stars with $N_\mathrm{free}=0$ for both sodium and magnesium. However, sodium is known to stabilise stacking less strongly, which we argue warrant sufficient flexibility to relax into a non-cubic crystalline phase.\\
Taken together, these results demonstrate the unique role played by structural flexibility in the crystallisation of amphiphilic DNA nanostars. Structural rigidity, along with directional binding, is considered as a necessary design feature for the crystallisation of conventional all-DNA building blocks interacting \emph{via} Watson-Crick base pairing or base stacking~\cite{Jones2015,Zhang2018,Zheng2009,Sha:2013aa}. The opposite is true for C-stars, which require a minimum degree of flexibility in order to relax into a configuration compatible with the long-range order, which as for all amphiphilic building blocks is imposed by symmetry and topology. The reversal of this basic design rule sets the present technology apart from other approaches, as it is much more straightforward to design a flexible DNA motif with a well defined topology than a stiff one with a precisely prescribed shape.

\ack
LDM and NJB acknowledge support from the EPSRC Programme Grant CAPITALS number EP/J017566/1. LDM acknowledges support from the Leverhulme Trust and the Isaac Newton Trust through an Early Career Fellowship (ECF-2015-494) and from the Royal Society through a University Research Fellowship (UF160152). RAB acknowledges support from the EPSRC CDT in Nanoscience and Nanotechnology (NanoDTC), grant number EP/L015978/1. WTK acknowledges support from an EPSRC DTP studentship. We acknowledge Diamond Light Source for provision of synchrotron beamtime (SM17271, SM16970) and we thank A. Smith and T. Snow for assistance in operating beamline I22, and N. Khunti for assistance with remote beamtime on beamline B21. We thank P. Cicuta for insightful discussions throughout this project and J. Doye for useful comments on the manuscript.\\

\section{Materials and methods}

\subsection{Oligonucleotide design and handling}

Nanostars with four-arms and a number of unpaired A bases at the central junction, $N_\mathrm{free}=0, 1, 4$ were designed using NUPACK.\cite{Nupack} Aside from the variation in $N_\mathrm{free}$, the sequences used for all nanostar designs were kept identical. Sequences are shown in Supplementary Material, Table S1. Non-functionalised strands were purified by the supplier (Integrated DNA Technologies) using standard desalting, while cholesterol-functionalised strands were purified by high-performance liquid chromatography. 
DNA samples, received lyophilised, were reconstituted in TE buffer (10\,mM Tris, 1\,mM EDTA, pH\,8.0, Sigma Aldrich), and their concentration was determined by UV absorbance at 260\,nm using a ThermoScientific Nanodrop 2000 UV-Vis spectrophotometer~\cite{Brady2017,Ryan2}.\\

\subsection{Small angle X-ray scattering: C-star aggregates}

Samples with C-star concentration of 5\,$\mu$M were prepared by mixing all required strands in stoichiometric quantities in TE buffer supplemented with either 300\,mM NaCl or 17\,mM MgCl$_{2}$. Samples were loaded and permanently sealed into flat borosilicate glass capillaries with internal section of $4$\,mm$\times 0.2$\,mm (CM Scientific) as detailed in Refs.~\cite{Brady2017,Ryan2}. The contents of 6 capillaries \emph{per} variant was extracted from the flat capillaries and concentrated by centrifugation and supernatant removal to a final C-star concentration of {\raise.17ex\hbox{$\scriptstyle\mathtt{\sim}$}} $100$\,$\mu$M, before being injected into borosilicate glass X-ray capillaries (diameter 1.6 mm) and left to sediment and form a macroscopic pellet at the bottom.\\
SAXS measurements on C-star aggregates were performed at the I22 beamline of the Diamond Light Source with wavelength of $\lambda = 1$ {\AA}, and beam size of approximately 300\,$\mu$m wide $\times$ 100\,$\mu$m high. The beam was scanned across the sample near the pellet region, collecting a single frame for each location with an exposure time of 100\,ms. Since the beam hits several randomly-oriented individual aggregates, the scattering patters collected from crystalline samples show the typical rings of powder diffraction (Fig.~\ref{figure4}). Patterns shown in Fig.~\ref{figure4} are the result of averaging over at least 20 different locations per sample. 2D patterns were radially averaged, before subtracting the background measured from a buffer-filled capillary. Furthermore an arbitraty logarithmic background was subtracted to account for the differences between individual glass capillaries.\\

\subsection{Small angle X-ray scattering: non-interacting nanostars}

Samples of non-interacting nanostars for solution SAXS studies,  lacking cholesterol modification, were prepared by mixing required oligonucleotides in TE buffer with 300\,mM NaCl or 17\,mM MgCl$_{2}$ to yield a final DNA concentration of 4~mg\,ml$^{-1}$ (approximately 50\,$\mu$M). Prepared mixtures were cooled from 95$^{\circ}$C to 20$^{\circ}$C at $-0.05^\circ$C\,min$^{-1}$ using a Techne\texttrademark~TC-512 thermocycler to enable nanostructure formation. Serial dilutions were prepared from this stock to yield three concentrations for each sample (4, 1.3, 0.4\,mg\,ml$^{-1}$ or approximately 50, 17, 6\,$\mu$M). Data presented here was collected at 1.3 mg\,ml$^{-1}$. SAXS measurements on freely diffusing nanostars were performed at the B21 beamline of the Diamond Light Source, with an accessible $q$ range of 0.0031 to 0.38~{\AA}$^{-1}$. Samples were loaded using an automatic sample changer into a quartz capillary enclosed in a vacuum chamber to reduce parasitic scattering. Temperature was held at 20$^{\circ}$C for all measurements. For all samples and corresponding buffers, a total of 10 frames each were collected and averaged. Samples showed no detectable radiation damage upon repeated exposure. Averaged buffer signal was subtracted from sample scattering prior to further data analyses. Guinier analysis was performed using a custom MATLAB script. Internal pair distribution functions, $P(r)$, were evaluated by the program GNOM\cite{Svergun:wi0087}.

\subsection{Simulations}

Coarse-grained molecular dynamic simulations were performed using oxDNA2~\cite{oxDNA,oxDNA2}, with an ionic strength equivalent to 300\,mM monovalent salt. All oligonucleotides comprising the non-cholesterolised nanostars were first generated with the experimental base-sequence, then forced to assemble into the sought motif using artificial harmonic traps under Virtual Move Monte Carlo~\cite{Whitelam:2007aa}. To avoid kinetic traps and speed up equilibration, at this stage base pairing was enabled only between nucleotides paired in the equilibrium structure.\\
Harmonic traps between complementary nucleotides were then removed, and the system was allowed to equilibrate under molecular dynamics (MD) for $10^{4}$ time steps, each with a duration of 3.03\,fs. The system's configuration was then sampled every $10^4$ time steps. The MD simulation uses an Andersen-like thermostat~\cite{Andersen:1980aa}, whereby the system evolves under Newtonain dynamics for 103 steps, followed by the velocities of some fraction of particles, as determined by the diffusion coefficient, being refreshed from a Maxwellian distribution at 290K. The diffusion coefficient was set to 2.5 simulation units, artificially high, to accelerated sampling of the arm distributions. For each of the three nanostar designs with $N_\mathrm{free}$, of 0, 1, and 4, 64 independent trajectories were run in parallel from the same initial configuration to improve statistics. Simulations were performed on the Darwin Supercomputer on the University of Cambridge High Performance Computing Service. \\
To compute simulated SAXS traces, the configuration of the system was sampled every $10^{6}$ time steps, and back-mapped to an all-atom representation using a previously verified approach~\cite{SANSFree}. Simulated X-ray scattering traces were generated using Crysol \cite{Crysol}, evaluating the scatting vector $q$ from $4\times10^{-3}$ to 0.4\,\AA$^{-1}$ with steps of $2\times10^{-4}$\,\AA$^{-1}$. The additional solvent electron density due to the presence of 300\,mM NaCl was found to have no significant impact on Crysol traces thus the solvent electron density was set to that of pure water, 0.334\,$e^-$\AA$^{-3}$. SAXS traces shown in Fig.~\ref{figure6} are the result of an average of 2000 independent configurations \emph{per} nanostar design. Intensities of experimental and simulated scattering traces were normalised by the first 10 $q$ points to allow for visual comparison. \\
   Probability distributions of the interarm angles as illustrated in Fig.~\ref{figure2} were determined for each nanostar design. Angles were determined though the normalised inner product between vectors connecting the centre of the nanostar to the end of each arm. The centre is defined as the mean position of the nucleotides nearest or at the junction: these are the 4 (nominally) paired nucleotides for the $N_{\text{free}}=0$ case, the 4 unpaired A nucleotides for the $N_{\text{free}}=1$ case, and the 16 unpaired A nucleotides for the $N_{\text{free}}=4$ case.\\
   Early into the simulation, the nanostar collapses into one of two conformers. Transitions between those conformers are rare for the $N_{\text{free}}=0, 1$ cases, complicating appropriate sampling of both conformers. This issue was solved here through dataset augmentation. The nanostar sequence within 2 nucleotides of the junction is symmetric under cyclic isomorphism: that is to say that under exchange of the arms of the nanostar, $(1,2,3,4) \rightarrow (4,1,2,3)$ \emph{etc.}, the sequences near the junction remain invariant. Nucleotide sequences only change three nucleotides away from the junction into the duplex of the arm, and consequently will not contribute to junction flexibility  (see Table S1 in SM). Therefore for each configuration of the four arms $(\vec{r_1},\vec{r_2},\vec{r_3},\vec{r_4})$, configurations were stored also for the three additional cyclic isomorphs $(\vec{r_4},\vec{r_1},\vec{r_2},\vec{r_3})$, imposing on the simulations the constraint of the symmetry of the junction.\\

\bibliography{CStarBib}{}
\bibliographystyle{unsrt}

\end{document}